\documentclass[manuscript,12pt,epsf]{aastex}

\usepackage{latexsym}

\shorttitle{X-ray Glint in the Cat's Eye}
\shortauthors{Chu et al.}

\begin{document}

\title{{\it Chandra} Reveals the X-ray Glint in the Cat's Eye}

\author{You-Hua Chu\altaffilmark{1}}
\author{Mart\'{\i}n A.\ Guerrero\altaffilmark{1}}
\author{Robert A. Gruendl\altaffilmark{1}}
\author{Rosa M. Williams\altaffilmark{2, 3}}
\author{James B. Kaler\altaffilmark{1}}
\altaffiltext{1}{Astronomy Department, University of Illinois, 
        1002 W. Green Street, Urbana, IL 61801, USA;
        chu@astro.uiuc.edu, mar@astro.uiuc.edu, gruendl@astro.uiuc.edu,
        kaler@astro.uiuc.edu}
\altaffiltext{2}{National Research Council Associate} 
\altaffiltext{3}{NASA's GSFC, code 662, Greenbelt, MD 20771, USA;
        rosanina@lhea1.gsfc.nasa.gov}

\begin{abstract}

We have obtained {\it Chandra} ACIS-S observations of NGC~6543, the
Cat's Eye Nebula.  The X-ray emission from NGC~6543 is clearly resolved 
into a point source at the central star and diffuse emission confined
within the central elliptical shell and two extensions along the major
axis.  Spectral analysis of the diffuse component shows that the
abundances of the X-ray-emitting gas are similar to those of the fast
(1,750 km~s$^{-1}$) stellar wind but not those of the nebula.  Furthermore,
the temperature of this gas is $\sim$1.7$\times$10$^6$ K, which is 100 
times lower than the expected post-shock temperature of the fast stellar
wind.  The combination of low temperature and wind abundances is puzzling.
The thermal pressure of this hot gas is about twice the pressure in 
the cool nebular shell; thus, the hot gas plays an essential role 
in the ongoing evolution of the nebula.

\end{abstract}


\newpage

\section{Introduction}

Optical emission-line images of planetary nebulae (PNs) reveal a fascinating
range of morphologies (e.g., the {\it Hubble Space Telescope} gallery of 
PNs\footnote{URL: http://oposite.stsci.edu/pubinfo/pr/97/pn/}),
indicating complex internal structures in the nebulae.  Among these PNs, 
NGC~6543, also known as the Cat's Eye Nebula, has perhaps the most interesting 
morphology.  As reported by \citet{Reed99}, the H$\alpha$ and [\ion{O}{3}] 
$\lambda$5007 line images of NGC~6543 are similar, showing an inner shell 
surrounded by an envelope with multiple, interlocking, semi-circular 
features.  The [\ion{N}{2}] $\lambda$6584 line image, on the other hand, shows 
bright clumps strung along arcs that appear to wrap around the 
envelope.  In addition, the [\ion{N}{2}] image shows two small curly features 
along the major axis and two linear jet-like features at 18$^\circ$ from the 
major axis.  (See Figure 1 of \citet{Reed99} for the [\ion{O}{3}] and 
[\ion{N}{2}] images of NGC~6543.)

What is the formation mechanism that has produced the complex nebular 
morphology of NGC~6543?  It has been suggested that PNs are formed 
by the current fast stellar wind sweeping up the circumstellar material 
lost previously via the slow AGB wind \citep[e.g.,][]{Kwok83,FBR90}.  
In this interacting-stellar-winds model, the physical structure of a PN 
is similar to that of a wind-blown bubble, as modeled by \citet{Weaver77}.
The PN will comprise a central cavity filled with shocked fast wind at 
temperatures of 10$^6-10^8$ K, a dense shell of swept-up AGB wind at 
10$^4$ K, and an outer envelope of unperturbed expanding AGB wind.  
This morphology is obviously too simple compared to the observed 
morphology of NGC~6543.  

NGC~6543 is a known X-ray source and was marginally resolved 
by {\it ROSAT} observations \citep{Kreysing92,GCG00}.  Diffuse X-ray emission 
implies the existence of hot gas.  It is thus important to resolve the 
diffuse X-ray emission and compare the distribution of hot gas to the 
location of the dense, cooler, nebular shell in order to understand the 
physical structure and formation mechanism of this nebula.

We have obtained {\it Chandra} observations of NGC~6543.  Its diffuse 
X-ray emission is clearly resolved into several components and shows 
excellent correspondence with some of the optical features.  In addition, 
a previously unknown point X-ray source is detected at the central star.
We have extracted and modeled the spectra of the diffuse X-ray emission
to derive the physical conditions of the hot gas.  The results and their
implications are reported in this Letter.  The analysis of the point 
source will be reported in a future paper.

\section{Observations}

NGC~6543 was observed with the Advanced CCD Imaging Spectrometer
(ACIS) on board the {\it Chandra X-ray Observatory} on 2000 May 10--11
for a total exposure time of 46.0 ks.  The nebula was placed at the
nominal aim point for the ACIS-S array on the back-illuminated (BI)
CCD chip S3.  The BI chip has a moderately higher sensitivity than the 
front-illuminated (FI) chips at energies below 1 keV.  Furthermore, 
the BI chip is not affected by the inadvertent radiation damage 
(occurred immediately after the deployment of {\it Chandra}) as were 
the FI chips.  The point spread function of the ACIS observation has a
half power radius (the radius encircling 50\% of the energy) 
$\sim$ 0\farcs5 at $\le$ 1 keV.  The energy resolution, $E/\Delta E$,
of the BI chip is $\sim 4.3$ at 0.5 keV and $\sim 9$ at 1.0 keV.

The observations were carried out at a CCD operating temperature of 
$-120^\circ$ C.  The background count rate is consistent with the 
quiescent background \citep{ACIS00}; therefore, no background ``flares" 
affected the observations and no time intervals needed to be removed.  
A total of 1,950$\pm$40 counts (background-subtracted) are detected 
from NGC~6543.

We received Level 1 and Level 2 processed data from the {\it Chandra}
Data Center.  The data reduction and analysis were performed using the
{\it Chandra} X-ray Center software CIAO V1.1.5 and HEASARC FTOOLS and
XSPEC V11.0.1 routines \citep{Arnaud96}.

\section{X-ray Emission from NGC~6543}

\subsection{Spatial Distribution}

The X-ray emission from NGC~6543 is clearly resolved into a point source 
and a diffuse component, as shown in Figure 1.   To describe the 
distribution of the X-ray emission, it is convenient to use optical features 
as points of reference; therefore, we plot the X-ray emission as contours
over an archival {\it HST} H$\alpha$ image.  The initial overlay
showed an offset of $\sim$1$''$ between the point X-ray source and the central 
star.  As this offset is within the range of combined pointing errors of 
{\it HST} and {\it Chandra}, we have shifted the X-ray image to register 
the point source at the central star.  The resultant X-ray contours overlaid
on the H$\alpha$ image are presented in Figure 1d.  The diffuse X-ray emission 
is well bounded by sharp, bright H$\alpha$ filaments.  This tight 
morphological correlation supports our choice of alignment between the
X-ray and H$\alpha$ images.

The X-ray-bounding H$\alpha$ filaments outline a 10$''\times8''$ 
elliptical shell with two $\sim3''$ outward extensions along the 
major axis (Figure 1c).  The northern extension appears to be comprised 
of two closed lobes, while the southern extension appears to consist of 
an incomplete lobe that is open toward the east.
The diffuse X-ray emission from the central elliptical shell is
limb-brightened and follows closely the inner wall of the nebular
shell in the H$\alpha$ image.  Diffuse X-ray emission is 
also present in the extensions.  In the northern extension, diffuse
X-ray emission fills each nebular lobe.  In the southern extension, 
diffuse X-ray emission appears to be bounded even though the nebular
lobe appears incomplete.

It is interesting to note that the western lobe in the northern 
extension and the eastern side of the southern extension appear to 
be radially aligned with the two linear, jet-like features best seen 
in Reed et al.'s (1999) [\ion{N}{2}] image at 17$''-23''$ from the 
central star.  This alignment may be naively used to suggest a physical
connection between the diffuse X-ray emission along this direction and 
the jet-like features; however, as we show in \S3.2, there is no 
spectral evidence supporting this hypothesis.

\subsection{Spectral Properties}

We have extracted spectra from the entire nebula and from three 
regions corresponding to the central elliptical shell, the northern 
extension, and the southern extension, respectively.  The X-ray spectrum 
of the central elliptical shell was extracted with the central star 
excised.  These spectra can be used to constrain the temperature, 
density, and abundances of the hot, X-ray-emitting gas, and to search 
for spatial variations in these physical conditions.
The extracted spectra are plotted in Figure 2.  In all cases the emission
peaks at 0.55 keV, and then drops abruptly to a faint plateau between
0.7 and 0.9 keV.  No significant emission is detected at energies greater 
than 1.0 keV.  Below 0.5 keV, the spectra of the central shell and the
southern extension drop off to nearly zero at 0.3 keV, while the spectrum 
of the northern extension levels and stays high at 0.3 keV.

The observed spectral shape depends on many physical parameters, 
including the temperature and chemical abundances of the X-ray-emitting 
gas, the intervening absorption, and the detector response.  We adopt
the thin plasma emission model of \citet{RS77}.
It is expected that the X-ray-emitting gas
should have a chemical composition consistent with either the 
fast stellar wind itself or a mixture of the fast wind and the nebular 
material.  The nebular abundances of He, C, N, O, and Ne have been 
reported to be 0.11, 2.3$\times$10$^{-4}$, 5.9$\times$10$^{-5}$, 
5.6$\times$10$^{-4}$, and 1.4$\times$10$^{-4}$ relative to hydrogen
by number, respectively \citep{AC83,Pwa84,MP89}.  Accordingly, we have 
adopted nebular abundances, relative to the solar values \citep{AG89}, of
1.13, 0.63, 0.53, 0.66, and 1.14 for He, C, N, O, and Ne, respectively, 
and 1.0 for the other elements.  For the abundances of the fast stellar 
wind, we have adopted He and N abundances 60 and 3 times the solar 
values, respectively, but kept the abundances of the other elements 
solar \citep{deKoter96}.  For the intervening absorption, we have assumed 
solar abundances and adopted absorption cross-sections from \citet{BM92}.

We first fit the observed spectrum of the diffuse X-rays from the entire 
nebula with models using the aforementioned nebular abundances, but with 
temperature, absorption column density, and the normalization factor as 
free parameters in the model.
No models with nebular abundances can match the observed spectral
shape.  More specifically, the models produce either too strong \ion{O}{7}
lines at $\sim$0.56 keV or too weak \ion{N}{6} lines at $\sim$0.43 keV.
The spectral fits can be improved only if the N/O ratio is raised
above the nebular value.  The stellar wind has an enhanced N/O ratio, 
roughly three times the nebular value; consequently, models using stellar 
wind abundances fit the observed spectral shape much more satisfactorily.

The best-fit model with stellar wind abundances has a plasma temperature 
of $T$ = 1.7$\times$10$^6$ K, an absorption column density of 
$N_{\rm H}$ = 8$\times$10$^{20}$~cm$^{-2}$, and a normalization factor 
of 7$\times$10$^{-5}$ cm$^{-5}$.  This best-fit model is overplotted on 
the spectrum of the entire nebula in Figure 2.  The reduced $\chi^2$ of 
the fits as a function of $N_{\rm H}$ and $kT$ is plotted in Figure 3.  
The 99\% confidence contour spans $N_{\rm H}$ = 
5.5--12$\times$10$^{20}$~cm$^{-2}$ and $T$ = 1.6--1.8$\times$10$^6$ K
(or $kT$ = 0.135--0.155 keV).  The observed (absorbed) X-ray flux is 
8$\times$10$^{-14}$ ergs cm$^{-2}$ s$^{-1}$; the unabsorbed X-ray flux 
is 8$\times$10$^{-13}$ ergs cm$^{-2}$ s$^{-1}$, and the X-ray luminosity 
is 1.0$\times$10$^{32}$ ergs s$^{-1}$ for a distance of 1 kpc 
\citep{CKS92,Reed99}.

We have also fit the spectra extracted from the central shell, and the
northern and southern extensions.  In these spectral fits we used
the same stellar wind abundances as those adopted in the fits for
the entire nebula.  These three spectra have fewer counts, so the 
best-fit plasma temperature and absorption column are not as well 
constrained as those for the entire nebula.  As indicated by the 
$\chi^2$ grid plots in Figure 3, the northern and southern extensions 
have very similar temperatures, but the northern extension has a 
smaller absorption column.  The temperature and absorption column of
the central shell are the least well constrained.  Its $\chi^2$ grid 
plot indicates that the gas in the central shell may be cooler, more 
absorbed, or both.  

The spectral fits suggest appreciable variation of absorption column 
across the 16$''$ extent of the X-ray emission region of NGC~6543.  
NGC~6543 is at a high Galactic latitude, $b = +29^\circ$, so the 
foreground interstellar absorption column is not likely to vary 
rapidly over the 16$''$ nebular extent.  The absorption column density 
through the nebular envelope of NGC~6543, on the other hand, is 
expected to increase from the north to the south because the major 
axis of the nebula is tilted with its north end toward us \citep{MS92}.
Furthermore, the nebular envelope is brighter and denser near the 
equatorial plane of the central shell, thus would produce higher 
absorption columns toward the central shell and the southern 
extension.  Therefore, the intervening absorption of the X-ray
emission from NGC~6543 occurs mostly within its cool nebular shell.

Finally, we note that the temperature appears uniform, no variations
greater than 50\% are detected.  We have extracted spatially-resolved
spectra for the two lobes in the northern extension (not shown here).
As described in \S3.1, the western lobe is radially aligned with a 
jet-like [\ion{N}{2}] feature, but both the eastern and western lobes 
display similar spectral shape, showing no evidence of additional 
dynamical heating in the western lobe.  Therefore, either the jet-like 
[\ion{N}{2}] feature is not a physically energetic phenomenon or the 
alignment is fortuitous.

\section{Discussion}

The physical properties of the hot gas in NGC~6543 can be compared
with those expected in the wind-blown bubble model of \citet{Weaver77}.
In this model, the bubble interior is filled by the adiabatically
shocked fast stellar wind; however, the heat conduction and mass 
evaporation across the interface between the hot interior and the 
cool nebular shell raise the density, lower the temperature, and 
alter the abundances of the hot gas near the interface.  
The limb-brightened X-ray morphology of NGC~6543 indicates that the 
hot gas near the interface is responsible for the X-ray emission.  
Furthermore, the temperature derived from X-ray spectral fits, 
$\sim$1.7$\times$10$^6$ K, is 100 times lower than the post-shock
temperature of NGC~6543's 1,750 km~s$^{-1}$ wind \citep{Peri89}, but
may be consistent with that expected near the interface.  
The location and temperature of NGC~6543's X-ray-emitting gas 
require that this hot gas contains a significant fraction of
nebular material.  In fact, if the mixing is adiabatic, the 
observed low temperature requires the X-ray-emitting gas to be
dominated by nebular material.  This is contrary to the results of
our spectral fits, which indicate stellar wind abundances!

As an independent test to the origin of NGC~6543's X-ray-emitting gas,
we compare the mass of the hot gas to the amount of fast stellar wind 
supplied during the lifetime of the nebula.  
The electron density of the X-ray-emitting gas can be derived from 
the normalization factor of the spectral fit and distance.
For a normalization factor of 7$\times$10$^{-5}$ cm$^{-5}$,
a distance of 1 kpc, and a He/H number ratio of 6,
the rms electron density is 50 $\epsilon^{1/2}$ cm$^{-3}$,
where $\epsilon$ is the volume filling factor.
The emitting volume, including the central shell and the extensions 
and taking into account of the 35$^\circ$ inclination of the major 
axis \citep{MS92}, is 2.5$\times$10$^{51}$ $\epsilon$ cm$^3$. 
The total mass of the X-ray-emitting gas is thus
2.5$\times$10$^{-4}$~$\epsilon^{1/2}$ M$_\odot$.  

The dynamical age of the Cat's Eye's central shell is $\sim$1,000 yr
\citep{MS92,Reed99}.  The stellar wind mass loss rate of the Cat's 
Eye's central star is uncertain, and ranges from 4$\times$10$^{-8}$ 
M$_\odot$ yr$^{-1}$ \citep{Peri89} to 3.2$\times$10$^{-7}$ 
M$_\odot$ yr$^{-1}$ \citep{Bianchi86}.  Within the dynamical age of the 
central shell, the total mass supplied by the stellar wind is 
4$\times$10$^{-5}$ M$_\odot$ to 3.2$\times$10$^{-4}$ M$_\odot$.
Comparisons between these values to the value derived from X-ray
observations indicate that the fast stellar wind can supply most
or all of the hot gas only if the high mass loss rate is adopted.

If the X-ray-emitting gas is comprised of largely shocked fast stellar
wind, the low temperature of the X-ray-emitting gas is puzzling. 
How does the hot shocked wind cool from 1.4$\times$10$^8$ K to 
1.7$\times$10$^6$ K within 1,000 yr?  It is possible that the fast 
stellar wind is not steady and its velocity was lower in the past, 
as suggested by \citet{Arnaudetal96} for BD~+30$^\circ$3639, 
another PN whose diffuse X-ray emission has been unambiguously 
resolved \citep{Kastner00}.  Future observations of more PNs with 
diffuse X-ray emission may help us solve this puzzle.

Finally, it is interesting to note that the {\it Chandra} image of
NGC~6543 shows the hot, X-ray-emitting gas to be confined within the 
central shell and the two extensions along its major axis, but not
associated with any of the intriguing optical features in the outer 
envelope.  The thermal pressure of the hot interior is about twice as 
high as the pressure in the nebular shell, for a nebular electron density 
of 4,000 cm$^{-3}$ and temperature of 7,000 K \citep{MP89}.  Clearly, 
the hot gas drives the expansion of the central shell, and governs
the ongoing evolution of NGC~6543, but is not responsible for the
complex optical features in the outer shell.

\acknowledgements
We thank the referee, Joel Kastner, for useful and constructive
suggestions to improve this paper.  This work is supported by the 
{\it Chandra} X-Ray Observatory Center Grant Number GO0-1004X.  
MAG is supported partially by the DGESIC of the Spanish Ministry 
of Education and Culture.

\clearpage

\begin{figure}

\caption{Image of NGC~6543 in (a) ACIS-S raw map; (b) ACIS-S adaptive filter
smoothed map; (c) HST WFPC2 image in H$\alpha$; (d) X-ray contours over the
H$\alpha$ image.  The contour levels are arbitrarily chosen to best follow
the surface brightness variations.}
\end{figure}

\begin{figure}
\epsscale{0.6}
\centerline{\plotone{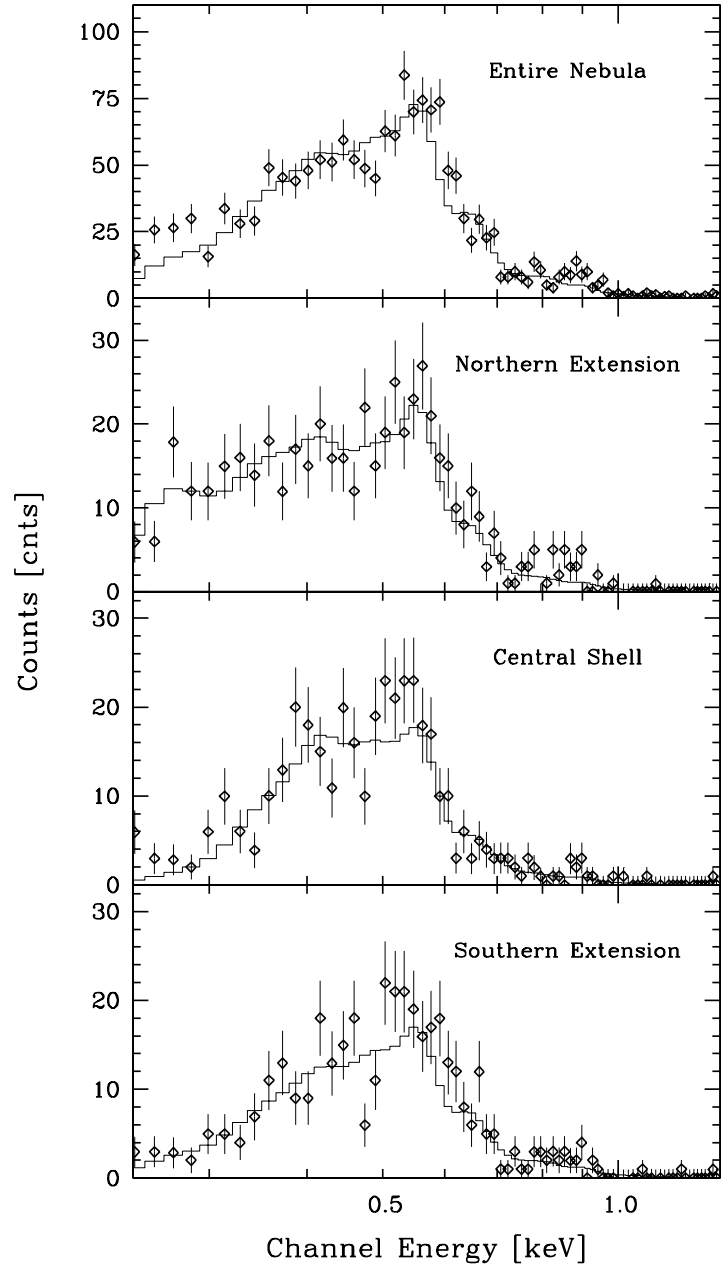}}
\caption{{\it Chandra} ACIS spectra with best-fit Raymond \& Smith models
overplotted.}
\end{figure}

\begin{figure}
\epsscale{0.6}
\centerline{\plotone{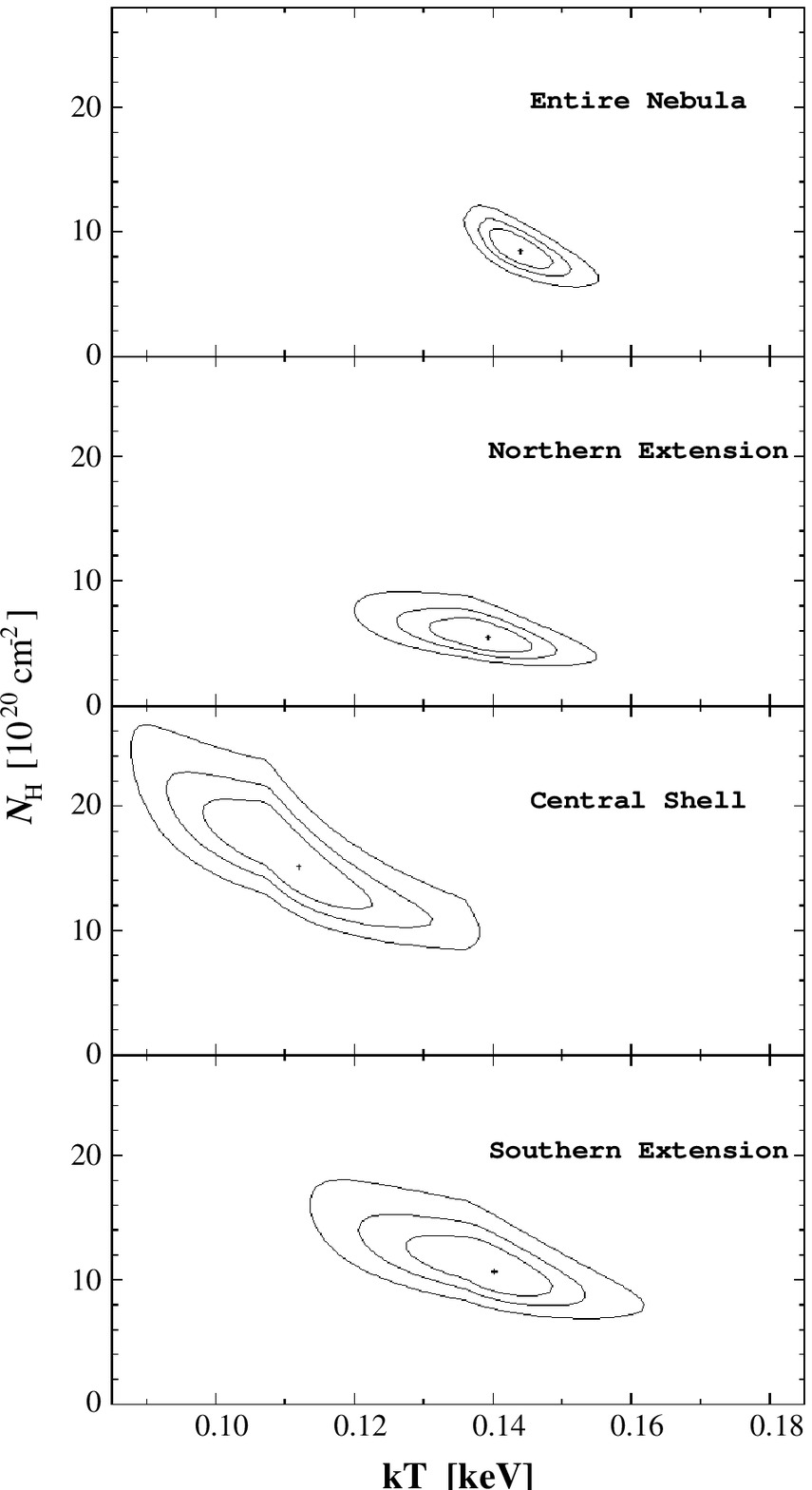}}
\caption{$\chi^2$ grid plots of the four spectral fits.  The contours
represent 68\%, 90\%, and 99\% confidence levels.}
\end{figure}


\begin{thebibliography}{}

\bibitem[ACIS Background Report (2000)]{ACIS00} ACIS Background Report 2000,
   prepared by Maxim Markevitch, 4/11/00, available at
   http://asc.harvard.edu/cal/Links/Acis/acis/Cal\_prods/bkgrnd/04\_11/bg110400.html

\bibitem[Aller \& Czyzak (1983)]{AC83} Aller, L.\ H., \& Czyzak, S.\ J.\ 
   1983, ApJS, 51, 211

\bibitem[Anders \& Grevesse (1989)]{AG89} Anders, E., \& Grevesse, N.\ 1989, 
   Geochimica et Cosmochimica Acta 53, 197

\bibitem[Arnaud (1996)]{Arnaud96} Arnaud, K.\ 1996, in ASP Conf. Ser. 101.
   Astronomical Data Analysis Software and Systems V, eds. G.\ Jacoby \& J.\ Barnes
   (San Francisco: ASP), 17

\bibitem[Arnaud, Borkowski, \& Harrington (1996)]{Arnaudetal96} Arnaud, K., 
   Borkowski, K.\ J., \& Harrington, J.\ P.\  1996, ApJ, 462, L75

\bibitem[Balucinska-Church \& McCammon (1992)]{BM92} Balucinska-Church, M.,
   \& McCammon, D.\ 1992, ApJ, 400, 699

\bibitem[Bianchi et al.\ (1986)]{Bianchi86} Bianchi, L., Cerrato, S., Grewing, M.\
   1986, A\&A, 169, 227

\bibitem[Cahn, Kaler, \& Stanghellini (1992)]{CKS92} Cahn, J.\ H., Kaler, 
   J.\ B., \& Stanghellini, L.\ 1992, ApJS, 94, 399

\bibitem[de Koter et al.\ (1996)]{deKoter96} de Koter, A., Heap, S.\ R., 
   Hubeny, I., \& Lanz, T.\ 1996, in ASP Conf. Ser. 96, Hydrogen deficient 
   stars, eds. C. S. Jeffery and U. Heber (San Francisco: ASP), 141

\bibitem[Frank, Balick, \& Riley (1990)]{FBR90} Frank, A., Balick, B., \& Riley, J.\ 
   1990, AJ, 100, 1903

\bibitem[Guerrero, Chu, \& Gruendl (2000)]{GCG00} Guerrero, M., Chu, Y.-H., \& 
   Gruendl, R. A.\ 2000, ApJS, 129, 295

\bibitem[Kastner et al.\ (2000)]{Kastner00} Kastner, J.\ H., Soker, N.,
  Vrtilek, S., Dgani, R.\ 2000, ApJ, 545, L57

\bibitem[Kreysing et al.\ (1992)]{Kreysing92} Kreysing, H.\ C., Diesch, C., 
   Zweigle, J., Staubert, R., Grewing, M., Hasinger, G.\ 1992, A\&A, 264, 623

\bibitem[Kwok (1983)]{Kwok83} Kwok, S.\ 1983, in IAU Symposium 103, Planetary Nebulae,
  ed. D.\ R.\ Flower (Dordrecht: Reidel), 293

\bibitem[Manchado \& Pottasch (1989)]{MP89} Manchado, A., \& Pottasch, S.\ R.\ 
    1989, A\&A, 222, 219

\bibitem[Miranda \& Solf (1992)]{MS92} Miranda, L.\ F., \& Solf, J. 1992, A\&A, 260, 397


\bibitem[Perinotto et al.\ (1989)]{Peri89} Perinotto, M., Cerruti-Sola, M.,
    Lamers, H.\ J.\ G.\ L.\ M.\ 1989, ApJ, 337, 382

\bibitem[Pwa et al.\ (1984)]{Pwa84} Pwa, T.\ H., Pottasch, S.\ R., \& Mo, 
   J.\ E.\ 1984, A\&A, 139, 1

\bibitem[Raymond \& Smith (1977)]{RS77} Raymond, J.\ C., \& Smith, B.\ W.\ 1977,
     ApJS, 35, 419

\bibitem[Reed et al.\ (1999)]{Reed99} Reed, D.\ S., et al.\ 1999, AJ, 118, 2430

\bibitem[Weaver et al.\ (1977)]{Weaver77} Weaver, R., McCray, R., Castor, J., Shapiro,
   P., \& Moore, R.\ 1977, ApJ, 218, 377



\end{thebibliography}
\end{document}